\documentstyle[12pt]{article}
\begin{document}

\begin{center}
{\Large{\bf The $\omega\rightarrow\infty$ limit of Brans-Dicke gravity revisited}}

\vspace{1,5cm}

Israel Quiros\\Dpto.Fisica. Universidad Central de Las Villas \\Santa Clara. CP: 54830 Villa Clara. Cuba\\Email: israel@mfc.esivc.colombus.cu
\end{center}

\vspace{2cm}

\begin{center}
\begin{abstract} 
The $\omega\rightarrow\infty$ limit of Brans-Dicke theory is studied with the help of the conformal transformation approach without resorting, however, to the conformal invariance property of this formalism, that is shown to be spurious.
\end{abstract}
\end{center}

\newpage

\section{Introduction}

Although Brans-Dicke(BD) theory[1,2] is a relatively old theory of gravitation some of its features are not well understood yet. In particular, the anomaly in the $\omega\longrightarrow\infty$ limit, associated to the vanishing of the matter stress-energy tensor[3], has been clarified only very recently by using the conformal invariance property of this theory[4].

In the present paper we retake the issue, and we shall show that BD theory is conformally invariant only in appearance so, in particular, the method followed in Ref.[4], though illuminating in general, is not justified. The anomaly in the $\omega\rightarrow\infty$ limit is treated here for the static, spherically symmetric case by using the conformal transformation approach as in [4] without resorting, however, to the spurious conformal invariance of the BD gravity.

It will honest saying that paper[4] served as motivation for our work.

\section{Conformal invariance of the BD gravity is spurious}

When we consider the purely gravitational sector of the BD theory given by the action

\begin{equation}
S_{BD}=\frac{1}{16\pi}\int d^4x \sqrt{-g} [\phi R-\frac{\omega}{\phi} g^{mn} \phi_{,n} \phi_{,m}]
\end{equation}

where $\phi$ is the scalar BD field, $\omega$ an adjustable coupling constant (a free parameter of the theory), under the conformal transformation:

\begin{equation}
{\tilde g_{ab}}=\phi^{2\alpha} g_{ab}
\end{equation}

and the field redefinition

\begin{equation}
\sigma=\phi^{1-2\alpha}
\end{equation}

with $\alpha\neq\frac{1}{2}$; the BD action (1) transforms into[4]:

\begin{equation}
S_{BD}=\frac{1}{16\pi}\int d^4x \sqrt{-{\tilde g}} [\sigma{\tilde R}-\frac{{\tilde \omega}}{\sigma}{\tilde g^{mn}} \sigma_{,n} \sigma_{,m}]
\end{equation}

where

\begin{equation}
{\tilde\omega}=\frac{\omega+6\alpha(\alpha-1)}{(1-2\alpha)^2}
\end{equation}

i.e., the gravitational part of the BD action(eq.(1)) is invariant in form under (2),(3) and (5). Hence the conclusion that BD theory is conformally invariant.

However, under transformations (2),(3) and (5), the Riemann geometry under which BD gravity rests[1], transforms into a more general 'Weyl-like' geometry. Actually, Riemann geometry is based upon the requirement that the scalar product of two vectors should remain unchanged under the transplantation law

\begin{equation}
d\xi^a=-\Gamma^a_{mn} dx^m \xi^n
\end{equation}

where $\xi^a$ are the components of an arbitrary vector ${\xi}$ in the given coordinate basis. Given another arbitrary vector $\zeta$ and a metric ${\bf g}$, the postulate of Riemannian geometry can be written as: 

\begin{equation}
dg(\xi,\zeta)=0
\end{equation}

where $g(\xi,\zeta)$ is the scalar product of the vectors $\xi$ and $\zeta$ in the metric ${\bf g}$. Under the transformation (2),(3), this postulate takes the form:

\begin{equation}
d\tilde g(\xi,\zeta)=\frac{2\alpha}{1-2\alpha}\tilde g(\xi,\zeta)\sigma^{-1}\sigma_{,n} dx^n
\end{equation}

which, in the metric ${\bf \tilde g}$, is the Weyl law of length(scalar product) transplantation with the vector potential $\psi_n\equiv\frac{2\alpha}{1-2\alpha}\sigma^{-1}\sigma_{,n}$ , such that the components of the antisymmetric tensor field ${\bf F}$ vanish[5]:

\begin{equation}
F_{mn}\equiv\psi_{n,m}-\psi_{m,n}=0
\end{equation}

This is the necessary and sufficient condition for a Weyl geometry to become Riemannian after an appropriate choice of the metric.

Law (8) results in that measuring rods will change length under bodily displacement unlike the situation we encounter in the usual Riemann geometry, where measuring rods keep their length unchanged when transported from one spacetime point to another. From this it follows that the free motion paths of material particles, being the geodesics of the Riemann geometry in the spacetime $(M,g_{ab},\phi)$ where $M$ is a 4D smooth manifold, in the spacetime $(M,\tilde g_{ab},\sigma)$ these are not geodesic paths anymore. In particular, for an uncharged, spinless mass point, we have in the metric with tilde $(d\tilde s^2=\tilde g_{mn} dx^m dx^n)$:

\begin{equation}
\frac{d^2 x^a}{d\tilde s^2}=-{\tilde \Gamma^a_{nm}}\frac{dx^n}{d\tilde s}\frac{dx^m}{d\tilde s}-\frac{\alpha}{1-2\alpha} \sigma^{-1}\sigma_{,n}[\frac{dx^n}{d\tilde s}\frac{dx^a}{d\tilde s}-\tilde g^{na}]
\end{equation}

instead of

\begin{equation}
\frac{d^2x^a}{ds^2}=-\Gamma^a_{nm} \frac{dx^n}{ds} \frac{dx^m}{ds}
\end{equation}

in the metric $\bf g$.

Then, although under the transformation (2),(3) and (5), the BD action (1) is invariant in form, the geometry under which the original theory rests, transforms from Riemann into 'Weyl-like'. Consequently, geodesic curves of the original Riemann geometry transform into non geodesic paths.

This leads to the conclusion that the purely gravitational sector of the BD gravity, that can be thought as the pair (BD action(1), test particle's free-motion path), is not really invariant under the conformal transformation (2), the field redefinition (3) and the parameter transformation (5).

Summing up: if $(M,g_{ab},\phi)$ is a BD spacetime then, in general, $(M,\tilde g_{ab},\sigma)$ it is not; so the conformal invariance of BD theory is only spurious.

\section{The static, spherically symmetric solution}

In this section we shall treat the situation with the anomaly in the $\omega\rightarrow\infty$ limit of BD gravity for the static, spherically symmetric case by using the conformal transformation approach as in [4], without resorting, however, to the conformal invariance property of the BD formalism, which has just been proved to be spurious.

When no ordinary matter is present, the BD field equations in the Jordan frame can be written as

\begin{equation}
G_{ab}=\frac{\omega}{\phi^2}[\phi_{,a}\phi_{,b}-\frac{1}{2}g_{ab} g^{nm}\phi_{,n}\phi^{,m}]+\frac{1}{\phi}\phi_{,a;b}
\end{equation}

\begin{equation}
\phi^{;n}_{;n}=0
\end{equation}

where $G_{ab}\equiv R_{ab}-\frac{1}{2} g_{ab} R$; meanwhile, the equations of free motion for an uncharged, spinless mass point are eqs.(11).

Under the conformal transformation:

\begin{equation}
\hat g_{ab}=\phi g_{ab}
\end{equation}

and the field redefinition

\begin{equation}
\psi=\ln \phi 
\end{equation}

eqs.(12),(13) and (11) transform into equations:

\begin{equation}
\hat G_{ab}=(\omega+\frac{3}{2})[\psi_{,a} \psi_{,b}-\frac{1}{2}\hat g_{ab} \hat g^{nm}\psi_{,n}\psi^{,m}]
\end{equation}

\begin{equation}
\psi^{;n}_{;n}=0
\end{equation}

and

\begin{equation}
\frac{d^2 x^a}{d\hat s^2}=-\hat{\Gamma}^a_{nm}\frac{dx^n}{d\hat s}\frac{dx^m}{d\hat s}+\frac{1}{2}\psi_{,n} [\frac{dx^n}{d\hat s}\frac{dx^a}{d\hat s}-\hat g^{na}]
\end{equation}

respectively, where covariant derivatives are now taken in respect to the metric ${\bf \hat g}$ $(d\hat s^2=\hat g_{mn} dx^m dx^n)$; i.e., the conformal transformation (14) recast the theory in the Einstein frame.

The static, spherically symmetric solution to eqs.(16) and (17) is well known[6]. In Schwarzschild coordinates it looks like ($d\Omega^2=d\theta^2+\sin^2 \theta d\varphi^2$):

\begin{equation}
{\it d}\hat{{\it s}}^2=-{\it f}^p{\it dt}^2+{\it f}^{-p}{\it dr}^2+{\it f}^{1-p}{\it r}^2{\it d}\Omega^2
\end{equation}

\begin{equation}
\psi=q\ln f
\end{equation}

with $f=1-\frac{2m}{pr}$. The parameters $p$ and $q$ are any two real constants such that:

\begin{equation}
p^2+(2\omega+3)q^2=1.
\end{equation}

We thus obtained a class of spacetimes labeled by two parameters, say $q$ and $\omega$:$(M,\hat g^{(q,\omega)}_{ab},\psi^{(q,\omega)})$.

The conclusions into which leads this solution for BD theory have been discussed in the literature[7].

Taking into account the relationship, $d\hat s^2=e^\psi ds^2$, one can obtain the corresponding solution in the Jordan frame:

\begin{equation}
ds^2=-f^{p-q} dt^2+f^{-p-q} dr^2+f^{1-p-q} r^2 d\Omega^2
\end{equation}

\begin{equation}
\phi=f^q
\end{equation}

leading to a class of spacetimes $(M,g^{(q,\omega)}_{ab},\phi^{(q,\omega)})$ labeled by the same parameters $q$ and $\omega$.

For convenience, the field equations (16) and (12) can now be rewritten as:

\begin{equation}
\hat G_{ab}=\frac{1}{2}(1-p^2)f^{-2}[f_{,a}f_{,b}-\frac{1}{2}\hat g_{ab}\hat g^{nm}f_{,n}f_{,m}]
\end{equation}

and

\begin{eqnarray}
G_{ab}&=&\frac{1}{2}(1-p^2-3q^2)f^{-2}[f_{,a}f_{,b}-\frac{1}{2} g_{ab} g^{nm}f_{,n}f_{,m}]+\nonumber\\&+&q[(q-1)f^{-2}f_{,a}f_{,b}+f^{-1}f_{;a;b}]
\end{eqnarray}

respectively. The curvature scalar take the forms:

\begin{equation}
\hat R=\frac{1}{2}(1-p^2)f^{-2}\hat g^{nm}f_{,n}f_{,m}
\end{equation}

in the metric ${\bf \hat g}$, and

\begin{equation}
R=\frac{1}{2}[1-p^2-3q^2)]f^{-2} g^{nm}f_{,n}f_{,m}
\end{equation}

in the metric $\bf g$, meanwhile the equation of motion (18) can now be written explicitly as:

\begin{equation}
\frac{d^2 x^a}{d\hat s^2}=-\hat{\Gamma}^a_{nm}\frac{dx^n}{d\hat s}\frac{dx^m}{d\hat s}+\frac{q}{2} f^{-1} f_{,n} [\frac{dx^n}{d\hat s}\frac{dx^a}{d\hat s}-\hat g^{na}]
\end{equation}

In order to study the $\omega\rightarrow\infty$ limit of BD gravity it is useful to rewrite eq.(21) as:

\begin{equation}
q=\pm\sqrt{\frac{1-p^2}{2\omega+3}}
\end{equation}

so in the $\omega\rightarrow\infty$ limit: $q\approx\pm\sqrt{\frac{1-p^2}{2\omega}}$.

For $\omega\rightarrow\infty$; $q\rightarrow 0$ independent on the value $p$ takes(with the constrain $p^2\leq 1$).

In this limit conformal transformation (14) approaches the identity transformation that leaves the theory in the same frame and, correspondingly, spacetime $(M,g^{(p)}_{ab})$ approaches $(M,\hat g^{(p)}_{ab})$ for a given $p$. In other words: in the $\omega\rightarrow\infty$ limit ($q\rightarrow 0$ limit), conformal transformation (14) maps the class of spacetimes $(M,g^{(p)}_{ab})$ into itself. This class of spacetimes is characterized by the line element (19) (we recall that $d\hat s\rightarrow ds$ in this case) and the case with $p=1$ recovers the Schwarzschild black hole solution of Einstein's general relativity, i.e., in this limit GR Schwarzschild solution is one of a class of solutions of the BD gravity.

The asymptotic behaviour of the BD scalar field can be rigorously derived in a way similar to that of Ref.[4]. Actually, in the $\omega\rightarrow\infty$ limit ($q\rightarrow 0$ limit) the BD scalar $\phi$ (eq.(23)) can be expanded in Taylor series in respect to the parameter $q$:

\begin{equation}
\phi\approx 1+q\ln f+\frac{1}{2} q^2(\ln f)^2+\cdots 
\end{equation}

Keeping members up to the first order in this expansion we obtain:

\begin{equation}
\phi\approx 1\pm\sqrt{\frac{1-p^2}{2\omega}}\ln f 
\end{equation}

which for solutions differing from that of GR($p^2\neq 1$) recovers the asymptotic behaviour $\phi\approx 1+O(\frac{1}{\sqrt{\omega}})$(see Ref.[4] and references therein).

\section{Conclusion}

We showed that BD theory is conformally invariant only in appearance, so the method in treating the $\omega\rightarrow\infty$ limit of this theory followed in [4] is not justified.

Instead, we gained more insight of the problem with the anomaly in the $\omega\rightarrow\infty$ limit of the BD gravity by using the conformal transformation approach without resorting, however, to the spurious property of conformal invariance of this formalism. For this purpose we have studied the static, spherically symmetric case of the BD theory without ordinary matter, and we showed that in the $\omega\rightarrow\infty$ limit, the GR Schwarzschild solution is one of a class of solutions given by the line element (19) (in this limit $d\hat s\rightarrow ds$) and laveled by the parameter $p$($p^2\leq 1$). Of these solutions, the one with $p=1$ recovers the Schwarzschild black hole solution of Einstein's theory.

Finally we provided a rigorous mathematical derivation of the asymptotic behaviour of the BD scalar field in the $\omega\rightarrow\infty$ limit (similar to that of Ref.[4]), for static, spherically symmetric solutions differing from that of GR ($p^2\neq 1$).

\newpage

\begin{center}
{\bf AKNOWLEDGEMENT}
\end{center}

We thank MES of Cuba for financing.

\begin{center}
{\bf References}
\end{center}
\begin{enumerate}
\item C.Brans, R.H.Dicke, Phys.Rev.{\bf 124}(1961)925
\item R.H.Dicke, Phys.Rev.{\bf 125}(1962)2163
\item N.Banerjee, S.Sen, Phys.Rev.{\bf D56}(1997)1334
\item V.Faraoni, Phys.Lett.{\bf A245}(1998)26
\item For a brief account on Weyl geometry see, for example: R.Adler, M.Bazin, M.Schiffer, 'Introduction to general relativity' (Mc Graw-Hill, 1965)401
\item A.Agnese, M.LaCamera, Phys.Rev.{\bf D31}(1985)1280
\item A.Tomimatsu in H.Sato, T.Nakamura (eds.), Gravitational collapse and relativity(World Scientific,1986)417
\end{enumerate}

\end{document}